\pdfoutput=1
\documentclass[namedreferences]{solarphysics}
\usepackage[pdfborder={0 0 0 },urlcolor=blue,breaklinks]{hyperref}
\ifx \doiurl    \undefined \def \doiurl#1{\href{http://dx.doi.org/#1}{\textsf{DOI}}}\fi
\ifx \adsurl    \undefined \def \adsurl#1{\href{http://adsabs.harvard.edu/abs/#1}{\textsf{ADS}}}\fi
\ifx \arxivurl  \undefined \def \arxivurl#1{\href{http://arxiv.org/abs/#1}{\textsf{arXiv}}}\fi
\usepackage[optionalrh]{spr-sola-addons} 
\usepackage{epsfig}          
\usepackage{graphicx}        
\usepackage{amssymb, amsthm,amsfonts,amsmath}        
\usepackage{amsmath}
\usepackage{amssymb}
\usepackage{color}           
\usepackage{breakurl}        
\usepackage{longtable}

\renewcommand{\vec}[1]{{\mathbfit #1}}

\newcommand{\astjp}{    {\it Aust J Phys}}
\newcommand{\naps}{    {\it Nature Phys. Sci.}} 
 
\newcommand{\lrs}{    {\it Living Rev. Solar Phys.}}

\newcommand{\apj}{    {\it Astrophys. J.}}
\newcommand{\apjl}{   {\it Astrophys. J. Lett.}}

\newcommand{\grl}{    {\it Geophys. Res. Lett.}}

\newcommand{\jastp}{  {\it J. Atmos. Solar-Terr. Phys.}} 
\newcommand{\jgr}{    {\it J. Geophys. Res.}}
\newcommand{\mnras}{  {\it Mon. Not. Roy. Astron. Soc.}}

\newcommand{\solphys}{{\it Solar Phys.}}
 
\newcommand{\ssr}{    {\it Space Sci. Rev.}} 
\chardef\us=`\_

\begin{document}

\begin{article}
\begin{opening}

\title{Intense Flare-CME Event of the Year 2015: Propagation and Interaction Effects between Sun and Earth's Orbit\\ {}}

\author[corref,email={johri@ncra.tifr.res.in}]{\inits{Abhishek}\fnm{Abhishek}~\lnm{Johri}}
\author[corref,email={mano@ncra.tifr.res.in}]{\inits{P.K.}\fnm{P.K.}~\lnm{Manoharan}}

\address[]{Radio Astronomy Centre, National Centre for Radio Astrophysics, Tata Institute of Fundamental Research, P.O. Box 8, Udhagamandalam(Ooty) 643001, India 
              Tel.: +91-4232-244965
              Fax: +91-4232-244900}
%
\runningauthor{Abhishek Johri and P.K. Manoharan}
\runningtitle{Intense Flare-CME Event of the Year 2015}

\begin{abstract}
In this paper, We report the interplanetary effects of a fast
coronal mass ejection (CME) associated with the intense X2.7 flare
 that occurred on 05 May 2015. The near-Sun
signatures of the CME at low-coronal heights $<$2 {R$_{\odot}$} are
obtained from the EUV images at 171 {\AA} and metric radio
observations. The intensity and duration of the CME-driven radio
bursts in the near-Sun and interplanetary medium indicate this CME
event to be an energetic one. The interplanetary scintillation
data, along with the low-frequency radio spectrum, played a
crucial role in understanding the radial evolution of the speed
and expansion of the CME in the inner heliosphere as well as its
interaction with a preceding slow CME. The estimation of
the speed of the CME at several points along the Sun to 1 AU shows
shows that i) the CME went through a rapid acceleration as
well as expansion up to a height of $\approx$6 {R$_{\odot}$}, and ii)
the CME continued to propagate at speed $\geq$800 kms$^{-1}$
between the Sun and 1 AU. These results show that the CME
likely overcame the drag exerted by the ambient/background
solar-wind with the support of its internal magnetic energy. When
the CME interacted with a slow preceding CME, the turbulence
level associated with the CME-driven disturbance increased
significantly.
\end{abstract}
\end{opening}

\section{Introduction}
\label{intro}
Coronal Mass Ejections (CMEs) are violent explosions in the atmosphere
of the Sun that carry massive magnetised plasma ({10}$^{12}$ $-${10}$^{13}$
kg) into the interplanetary medium. Their velocities range from 10 to 2000
kms$^{-1}$. CMEs have been identified as the cause of major geo-magnetic
storms and many studies have been made to understand their times of arrival as well as effects at 1 AU. Since CMEs can evolve considerably in speed
and shape on their way from the Sun to 1 AU, the propagation effects of the CMEs
in interplanetary space become essential in predicting their arrival at
1 AU. However, the relative importance of evolutionary processes can differ
from one event to the other and it depends on i) the initial energy of
the eruption of the CME, ii) solar-wind conditions along its path of 
propagation, and iii) the physical characteristics of the CME
 (\textit{e.g.} Gosling \textit{et al.}, 1998; Gopalswamy \textit{et al}., 2001a; Manoharan \textit{et al}.,
2001).
Moreover, since a CME phenomenon is three-dimensional and ambient solar
wind surrounding the CME propagation is often highly structured in space
and time, the CME-driven disturbances in the solar-wind
can be more complex. Therefore, for a better understanding of the effects
of the propagation of a CME, multi-point observations between Sun and 1 AU are
required.

In this article we report the propagation effects of a fast CME
associated with the intense X2.7 flare  that occurred
on 05 May 2015 at N15E75. In determining the speed of a CME, the
effects of projection can be crucial. Since the above event originated
close to the east limb of the Sun, projection effects may not be
significant. In this multi-wavelength analysis, the near-Sun signatures
of the eruption have been obtained from the
Atmospheric Imaging Array (AIA: Pesnell, W.D., Thompson, B.J., and Chamberlin, P.C., 2012) onboard the Solar {Dynamics} Observatory (SDO) at 171 {\AA}. The white-light
images from the Large Angle Spectroscopic Coronagraph (LASCO: Brueckner, G.E., Howard, R.A., and Koomen, M.J., 1995) onboard the SOlar and Heliospheric Observatory (SOHO),
combined with the radio measurements from Hiraiso Radio Spectrograph (HiRAS) and
WAVES radio experiment of \textit{Wind} spacecraft (Bougeret, J.L., Kaiser, M.L., and Kellogg, P.J., 1995) spectra, have been employed to infer the speed
and energetics of the CME at heights of $\leq$20 {R$_{\odot}$} (where, 1 solar radius ({{R$_{\odot}$}}) = 6.96 $\times$ {10}$^{5}$ km). Midway between
Sun and 1 AU, the CME-driven disturbances as well as its interaction
characteristics with a preceding slow CME have been inferred from
interplanetary scintillation observations made with the Ooty Radio
Telescope (ORT: Swarup \textit{et al}., 1971). 


\section{Observations of AR\#2339}
\label{obs}
The current solar cycle seems to be running low in activity in comparison 
with the recent previous cycles (\textit{e.g.} Hathway, 2010; Manoharan, 2012). However, a couple of active regions of the
current cycle were exceptionally active and produced a large number of flare
events and associated CMEs. One such active region,
AR2339, became active from its appearance at the east limb of the Sun on 
04 May 2015 and continued until it rotated to the back of the Sun on 17 May 
2015. It had an overall magnetic configuration of $\beta\gamma$ and grew to 
a fairly large area of $\approx$840 millionths on 10 May 2015. 
AR2339 produced 
more than 60 X-ray flares of C-class intensity  or more, during its passage from 
east to west limbs of the Sun. It is to be noted that the number of optical flares 
produced by this active region outnumbered the X-ray events. However, the rate
of occurrence of flares gradually declined after its crossing of the central 
meridian of the Sun. In this study, we consider the near-Sun, interplanetary propagation effects of a fast CME that occurred on 05 May 2015 at this active region as well as its interaction effects in the Sun--1 AU distance.

\subsection{X-ray and EUV Observations}
\label{ndo}
\begin{figure*}
\begin{center}

\includegraphics[scale=.53]{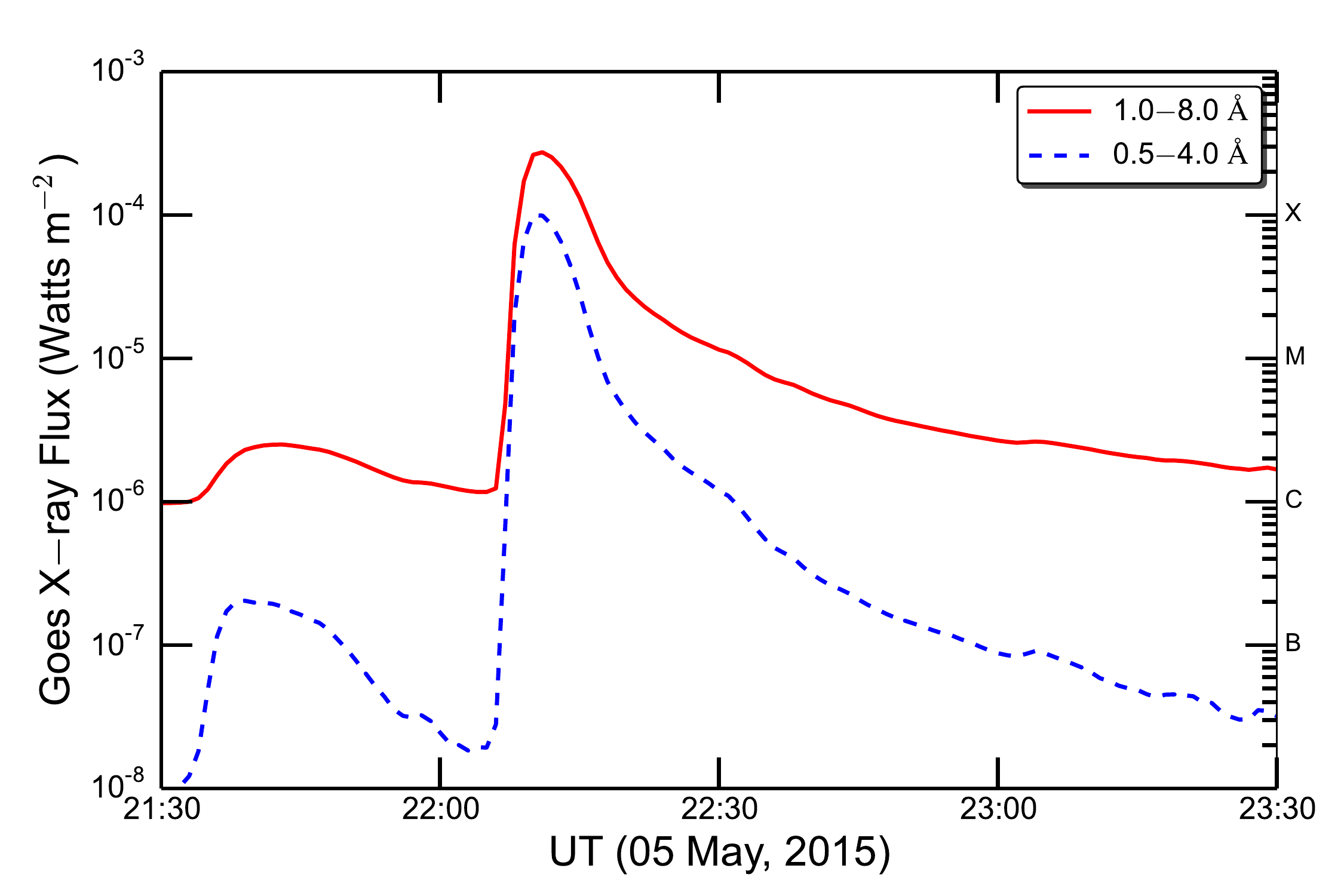}
\caption[caption ]{Soft X-ray light curves obtained from GOES-15 spacecraft for the period of 21:30--23:30 UT. The continuous   and dashed curves represent, respectively, 1--8 and 0.5--4 {\AA} channels.}
\label{xray-sdo-co}
\end{center}
\end{figure*}

On 05 May 2015, when the AR2339 was located close to the east limb of 
the Sun at N15E75, the onset of the X2.7-class flare  was observed at about 22:05 UT
in the wavelength bands of 1--8 {\AA} and 0.5--4 {\AA} {by the X-ray Sensor (XRS)\footnote{www.ssec.wisc.edu} onboard the GOES-15 spacecraft}. 
Figure \ref{xray-sdo-co} displays the X-ray flux profiles of the 
flare event during the period 21:30--23:30 UT in the above two wavelength bands.
The fluxes in these channels started to rise at 22:07 UT and reached a maximum value within about four minutes and returned to  the 
background value after about two hours. It is also to be noted that the harder part (0.5--4 
{\AA}) of the X-ray flux increased by more than about three orders of 
magnitude in comparison with its mean background value. However, the 
soft (1--8 {\AA}) X-ray spectrum increased only about 
two orders with respect to its background. It is likely that the heating 
at the flare site was efficient to produce the hard part of the X-ray 
spectrum.

\begin{figure*}
\begin{center}
\includegraphics[width=3.8 cm,angle=90,clip=true]{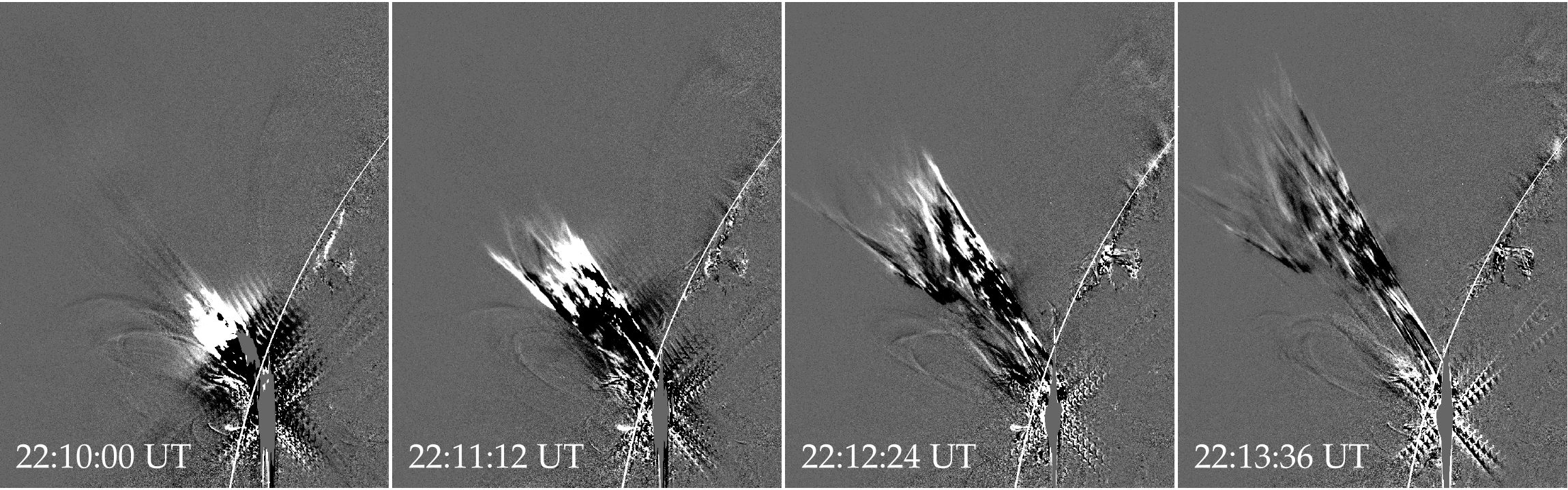}
\caption[caption ]{Sequence of running difference partial-frame images from Solar {Dynamics} Observatory at 171 {\AA}. White arc in the figure represents the  solar limb.}
\label{sdo-diff}
\end{center}
\end{figure*}

\begin{figure*}
\begin{center}
\includegraphics[scale=.36]{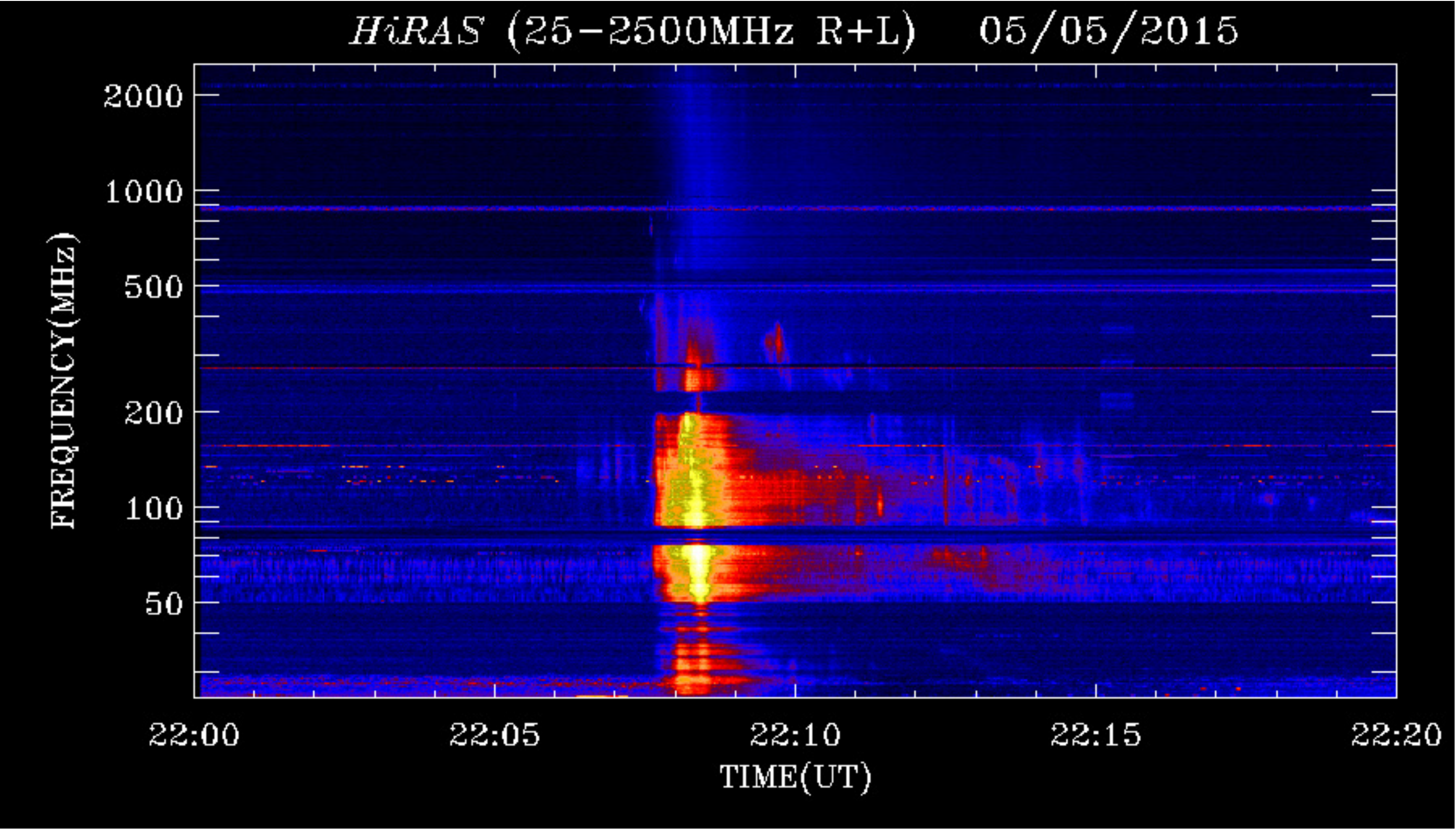} \\ {{\Large (a)}}
\includegraphics[scale=.28,angle=0]{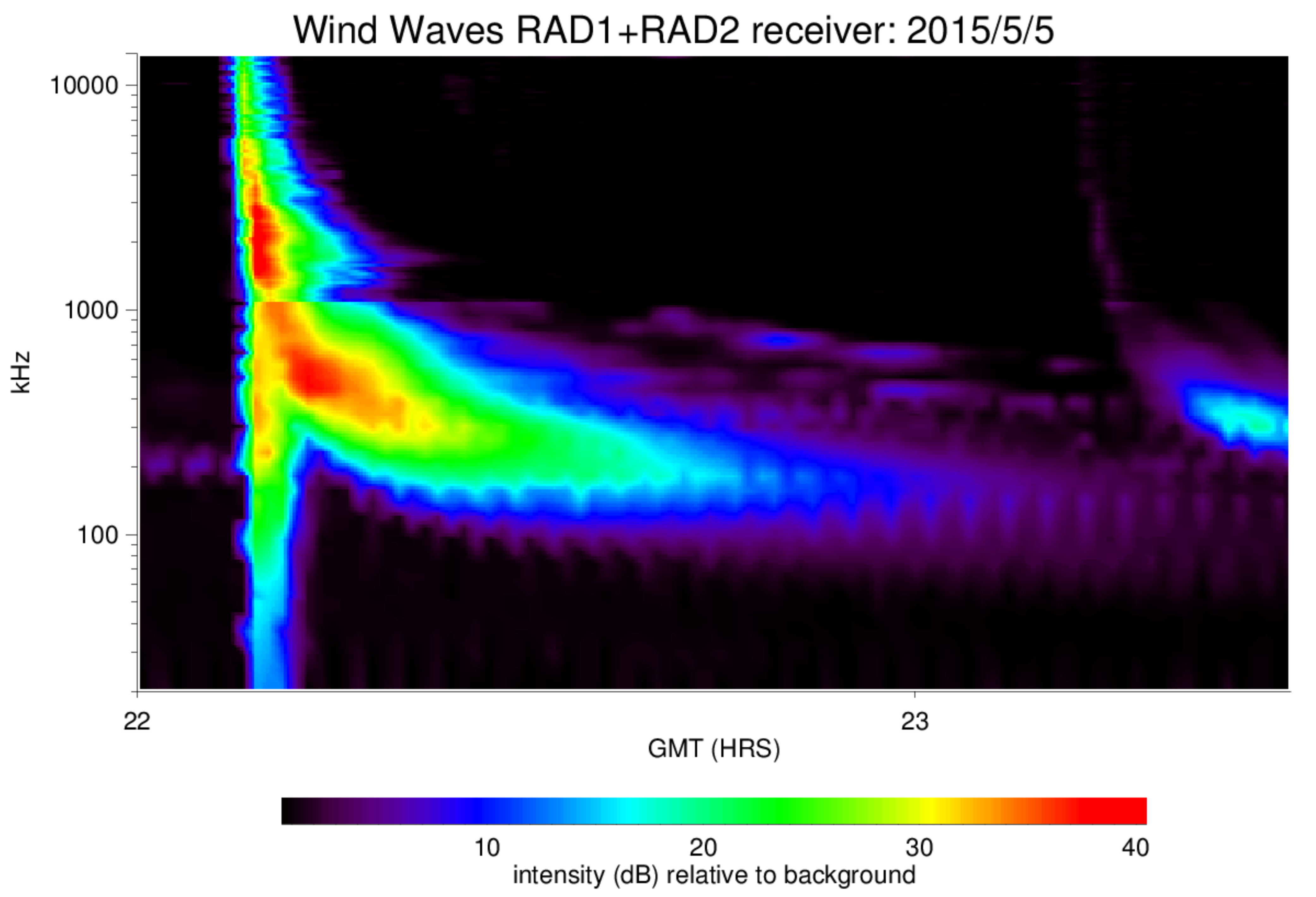}\\ {{\Large(b)}}
\includegraphics[scale=.19]{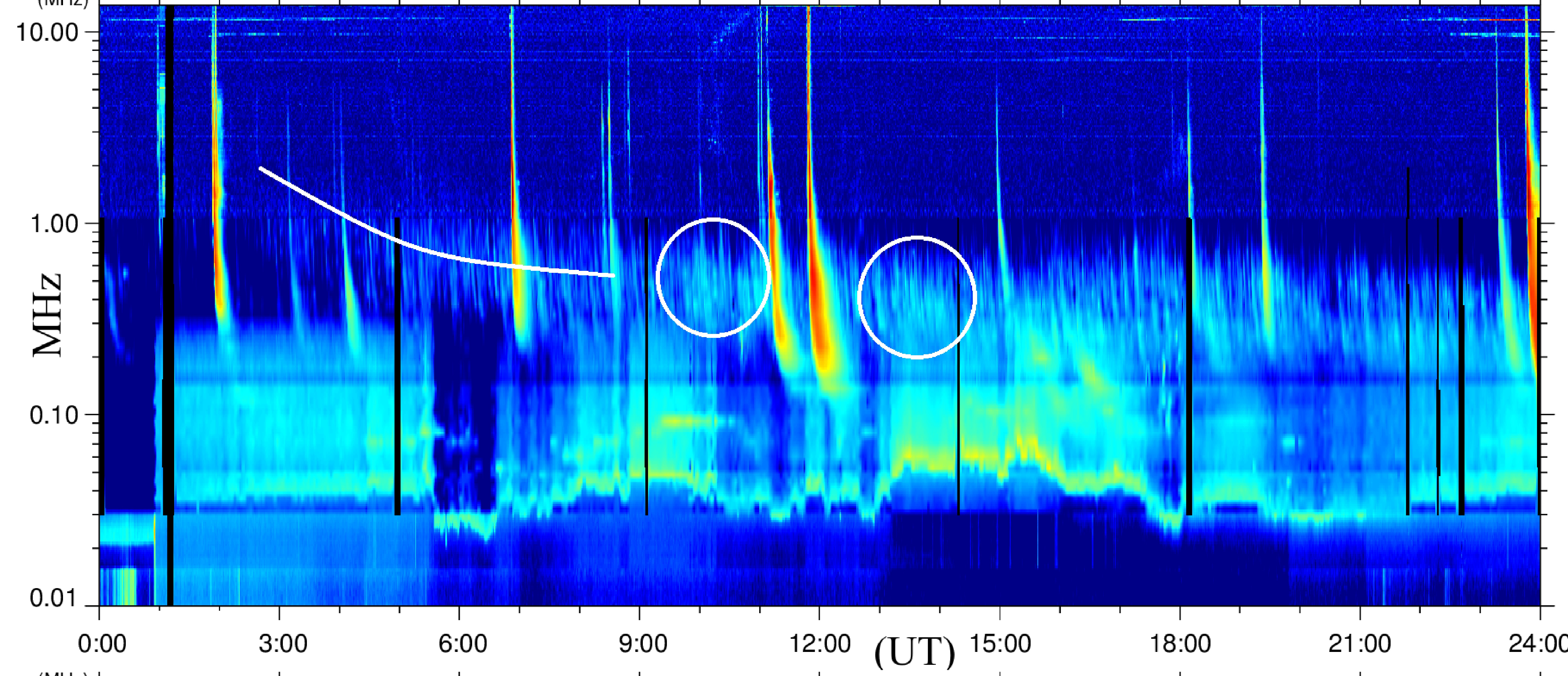}\\  {{\Large (c)}}
\caption[caption ]{Solar-radio-dynamic spectra  obtained from (a) HiRAS, (b) \textit{Wind}/WAVES on 05 May 2015, and (c) \textit{Wind}/WAVES on 06 May 2015. In panel (c) low-frequency drift is indicated by a continuous white line. Circles indicate the enhanced low-frequency emission at the time of CME--CME interaction (refer to Section \ref{cme-cme-int}).}
\label{comb-spec}
\end{center}
\end{figure*}

\subsubsection{EUV Near-Sun Images}
\label{euv}
AIA recorded the eruption of the CME associated with the X2.7 flare event, 
in a wideband of extreme ultraviolet (EUV) emission. Before the onset 
of the X-ray flare, AIA images showed the destabilisation of the filament
at the flare site and the eruption of the filament was seen at $\approx$22:07 UT. 
The eruption was also associated with the enhanced emission of X-ray (refer to Figure \ref{xray-sdo-co})
as well as EUV emission over a band of wavelengths 94--1700 {\AA}. 
At the rising phase of the flare, the AIA images showed the radial motion 
of mass and flux rope along the position angle of $\approx$50$^{\circ}$. 
Figure \ref{sdo-diff} shows the running-difference partial frame images 
from AIA observed in the 171 {\AA} band. At the time of 
maximum of the X-ray profile, $\approx$22:11 UT, the eruption seen in EUV 
reached a height of about 1.5 {R$_{\odot}$}, with respect to the centre of the 
Sun. The 
projected speed of the EUV eruption in the plane of the sky along its 
propagation direction is $\approx$1300 kms$^{-1}$. Since the location of
the CME origin is close to the limb of the Sun, the projection effect
is expected to be minimal and the sky-plane speed represents the actual
speed of the CME. However, the rate of change of width of the 
eruption in the direction perpendicular to its radial direction was 
significantly less and the speed of the lateral expansion was typically 
$\approx$300 kms$^{-1}$. {For example, in the AIA field of view, the angular width of
the eruption typically increased from $\approx$3$^{\circ}$ to $\approx$7$^{\circ}$ in the position angle, which determined the
typical lateral expansion. These rates correspond to a radial expansion of about four times faster than the
lateral expansion. The above east-limb eruption was also recorded by the Sun Watcher with Active Pixel System detector and Image Processing (SWAP) telescope at 174 {\AA} onboard the Project for On-Board Autonomy-2  spacecraft (PROBA-2: Halain \textit{et al.}, 2013; Seaton \textit{et al}., 2013). The findings obtained from AIA are consistent with the SWAP measurements.} The expansion of the CME and its associated filament 
eruption could be tracked up to about 1.5 {R$_{\odot}$}.
  
  \subsubsection{Radio Signatures of the CME}
  \label{radio}
  The radio signatures associated with the flare/CME event were observed 
over a wide range of frequencies: i) HiRAS 
provided the radio manifestations of the CME in a frequency range of 30  
to 2500 MHz, and ii) \textit{Wind}/WAVES covered  frequencies below 14 MHz. The
above spectra are useful to infer the radio signatures of the CME in the 
{solar-height} range of about 0.1 to $\geq$30 {R$_{\odot}$}. Figure \ref{comb-spec} shows 
the radio spectra observed on 05 and 06 May 2015 by the HiRAS solar radio 
spectrograph and \textit{Wind}/WAVES space mission. At the time of the CME onset, 22:07 UT, the opening of bundle 
of magnetic field lines were revealed by a group of intense Type-III bursts 
at frequencies below 500 MHz (refer to Figure \ref{comb-spec}a). The well-organised collimated eruption seen with 
AIA (refer to Figure \ref{sdo-diff}) is also consistent with the propagation of the CME along the open-field lines (\textit{i.e.} magnetic funnel type of structure) formed above the 
system of loops at the active region. In the HiRAS spectrum (Figure \ref{comb-spec}a), a broad-band 
diffused emission was also observed at high frequencies ($\approx$700--1800 MHz) 
around 22:07 UT and it is  likely associated with the on-going small-scale 
reconnections at the flare site at low coronal heights. 
  
  The fast eruption seen in the AIA EUV images has caused a shock, 
which is consistent with the intense fast drifting metric Type-II radio 
burst observed in the frequency range of $\approx$50--80 MHz during 22:12:00 to 
22:13:30 UT (refer to AIA images in Figure \ref{sdo-diff} and the HiRAS 
spectrum shown in Figure \ref{comb-spec}a).  In the above time range, a fast 
drifting Type-II was also independently observed by the 
Culgoora Radio Spectrograph in the 
frequency range of $\approx$25--40 MHz.  These spectra from Culgoora and HiRAS 
observatories confirm, respectively, the fundamental and second harmonic emissions 
of the Type-II burst (\textit{e.g.} Wild, J.P., Murray, J.D., and Rowe, W.C., 1954). This harmonic emission was also seen weakly in the HiRAS spectrum. The frequency coverage of the Type-II shows the 
typical height of the shock between 1.6 and 1.8 {R$_{\odot}$}, with respect to 
the centre of the Sun (\textit{e.g.} Pohjolainen \textit{et al.,} 2007). {At these heights, the overall drift rate of the Type-II burst suggests a shock speed of  $\approx$1500 kms$^{-1}$. 
In fact, the shock speed associated with an eruption is always higher than the average speed of
the eruption. In the present case, the observed Type-II speed and the radial speed of the eruption
seen in the AIA images (as well as from PROBA-2/SWAP images) at nearly similar solar heights
are consistent.}

Another important point is that the intense high-frequency Type-III 
bursts observed in the HiRAS spectrograph, continued to the interplanetary 
medium as shown by the \textit{Wind}/WAVES spectrum (Figure \ref{comb-spec}b), and manifested as an intense  
long-duration burst, lasting for more than 90 minutes, at frequencies 
below 1 MHz. This confirms that at the flare site a copious amount of 
energetic electrons were produced and pumped into  interplanetary
space (as the flare-associated eruption led to the opening of 
the magnetic-field lines), which extended to larger heights in the 
interplanetary medium. Such low-frequency long-duration Type-III radio 
bursts, when associated with Type-II events have shown statistical 
association with solar energetic particle (SEP) events \citep{
macdowall2003, gopal2010}. Thus, if the origin of the 
event under study were favourably located with respect to the Earth, 
a particle event would most probably have been observed at the Earth.
Additionally, the Type-II burst signatures were also observed in
the low-frequency part of the \textit{Wind}/WAVES spectrum, at frequencies below
1 MHz (however, the Type-II feature was not visible in the 1--14 MHz 
part of the spectrum). A broadband drifting feature seen below 1 MHz 
started at about 02:00 UT (Figure \ref{comb-spec}c) and by that time the CME had crossed a {solar 
height} of $\approx$18 {R$_{\odot}$} in the LASCO-C3 field of view. 
{The wide frequency range of Type-III bursts in the near-Sun as well as in the interplanetary
medium (including fast Type-II burst) suggests that the CME was involved with a significant high
energy \citep{macdowall2003, gopal2010}.}

As the CME moved away from the active region (before and after
the onset of the Type-II burst), intense continuum emissions were 
also recorded by both Culgoora and HiRAS spectrographs, in the frequency 
range of 50--200 MHz. Additionally, within the above continuum, there 
were several vertical structures. After the end of the Type-II 
burst, the continuum was observed at frequencies below 70 MHz. However, 
the intensity of the above radio continuum was much weaker than the 
Type-III and Type-II radio bursts. The signatures of the broadband 
continuum at times before the onset of the Type-II burst suggests 
that the emission originating from the system of rising loops resulted 
at the time of the CME eruption. The weak low-frequency continuum emission 
during the Type-II and up to $\approx$22:15 UT is likely to be associated 
with the expansion of the filament and the energetic electrons trapped 
within the propagating CME structure \citep{gosling1998,mano2003}.

\subsection{LASCO White-Light Images and CME Speed}
\label{lasco}
\begin{figure*}
\begin{center}
\includegraphics[width=12 cm,clip=true]{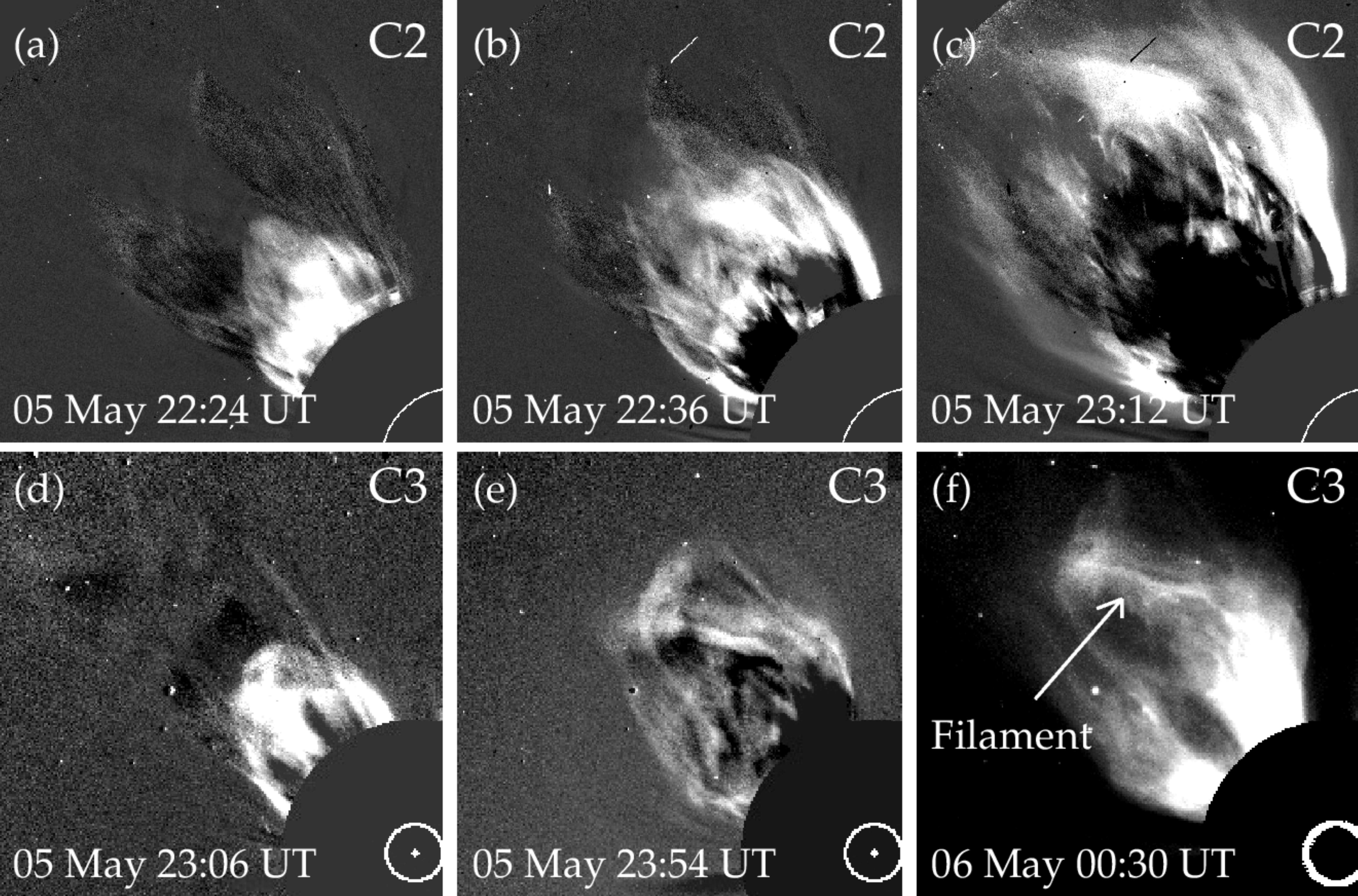}
\caption[caption ] {Running-difference images from LASCO-C2 and -C3 coronagraphs as indicated, observed on 05 May 2015. Panel f shows the raw image (without running subtraction) and filament  is indicated by an arrow. The limb of the Sun is shown by white line. }
\label{lasco-diff}
\end{center}
\end{figure*}

The onset of the fast moving CME was observed at 22:24 UT in the LASCO-C2
field of view at 4.3 {R$_{\odot}$}. Figure \ref{lasco-diff} shows the sequence 
of running difference white-light images from C2 and C3 coronagraphs 
of LASCO. 
In these ``position angle-distance" images, the radial evolution of 
the propagating CME is evidently seen in C2 (Figure \ref{lasco-diff}a,c) and C3 (Figure \ref{lasco-diff}d,f) fields of view
and the bright embedded filament structure can also be clearly 
identified (in Figure \ref{lasco-diff}f). Moreover, as observed in Figure \ref{lasco-diff}a, 
when the CME onset occurred in the C2 coronagraph, the lateral width 
of the CME (\textit{i.e.} in the perpendicular orientation to that of the
propagation direction along the position angle of $\approx$50$^{\circ}$) 
was more than 2 {R$_{\odot}$}. This reveals that the overall size of the CME 
has gone through a rather rapid expansion and also CME has been 
accelerated between AIA and LASCO fields of view. The rate of lateral 
expansion of the CME corresponds to a speed of  $\approx$900 kms$^{-1}$.  It is likely that the CME material, the ejecta as well as the magnetic 
field, have gone through the process of pressure balance with the 
surrounding solar-wind. For example, a CME can expand, if it is 
ejected from the Sun with a speed higher than the ambient solar-wind
causing a speed gradient. The ambient solar-wind estimates, obtained 
from the interplanetary scintillation measurements made with 
the ORT at a distance mid way between the Sun 
and Earth and from the in-situ data at 1 AU, indicate a low-speed 
heliosphere of about 300--350 kms$^{-1}$ in the space ahead of the CME.
It is likely that in order to maintain the  pressure balance between
the CME and the ambient solar-wind, the CME has gone through the
sudden expansion.

\begin{figure*}
\begin{center}
\includegraphics[width=12 cm,angle=180,clip=true]{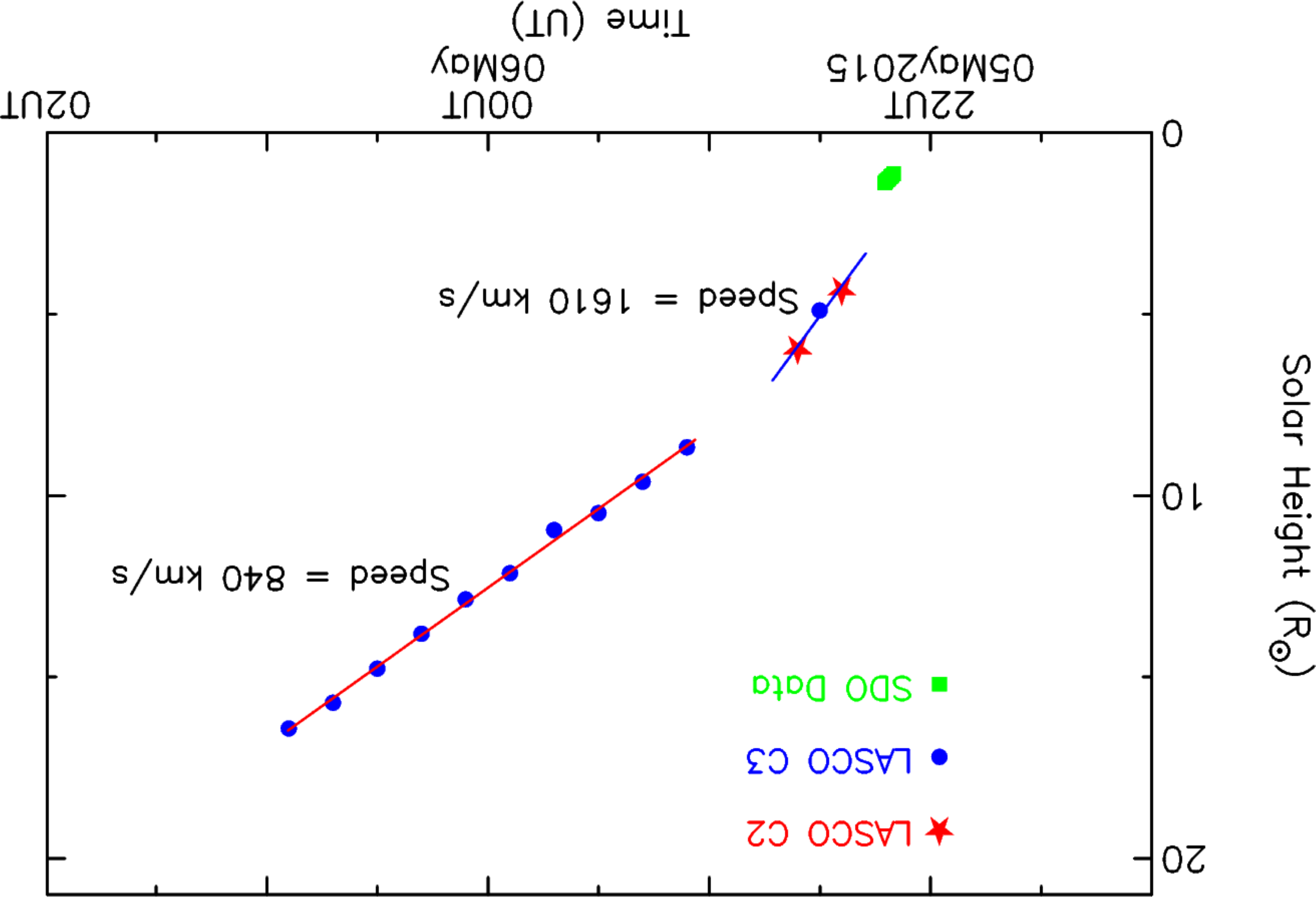}
\caption[caption]{CME height--time plot made using AIA and LASCO (C2 and C3) images. The speeds at respected segments are indicated.}
\label{lasco-ht-plot}
\end{center}
\end{figure*}

The quick acceleration of the CME along its radial direction has also 
been shown by the height--time plot obtained from the AIA EUV and LASCO white-light 
images. Figure \ref{lasco-ht-plot} shows the height--time plot of the CME and its derived 
speeds in the LASCO field of view. The height--time plot has been restricted 
to C2/C3 white-light images having good contrast and the low signal-to-noise 
ratio images are not considered. This plot also includes the height--time 
measurements obtained from the AIA images observed at 171 {\AA}. An important 
point revealed in this plot  is that when 
the initial three measurements of the LASCO images are considered (\textit{i.e.} 
two images from C2 and one image from C3 coronagraphs), the speed of the CME is 
rather high, around 1600 kms$^{-1}$. As stated earlier, since the originating 
location of the CME on the Sun is close to the east limb, the projection 
effect on the speed estimate obtained from the plane of sky images is 
expected to be less significant. The speed obtained from the initial images 
of LASCO is consistent with the speed of the shock (1500 kms$^{-1}$)  estimated from the Type-II 
radio burst at distances between AIA and LASCO fields of view (refer to Section \ref{radio}). In fact, the CME 
has thus gone through a steady acceleration in the near-Sun region between 
1.5 and 6 {R$_{\odot}$} (Figure 5). At greater heights between 8 and 
20 {R$_{\odot}$} in the C3 field of view, the speed of the CME however remained  
nearly constant at about 840 kms$^{-1}$, which was also well above the 
ambient solar-wind speed. 

\section{CME--CME Interaction and Interplanetary Scintillation Measurements}
\label{cme-cme-int}
\begin{figure*}
\begin{center}
\includegraphics[scale=0.18,angle=0]{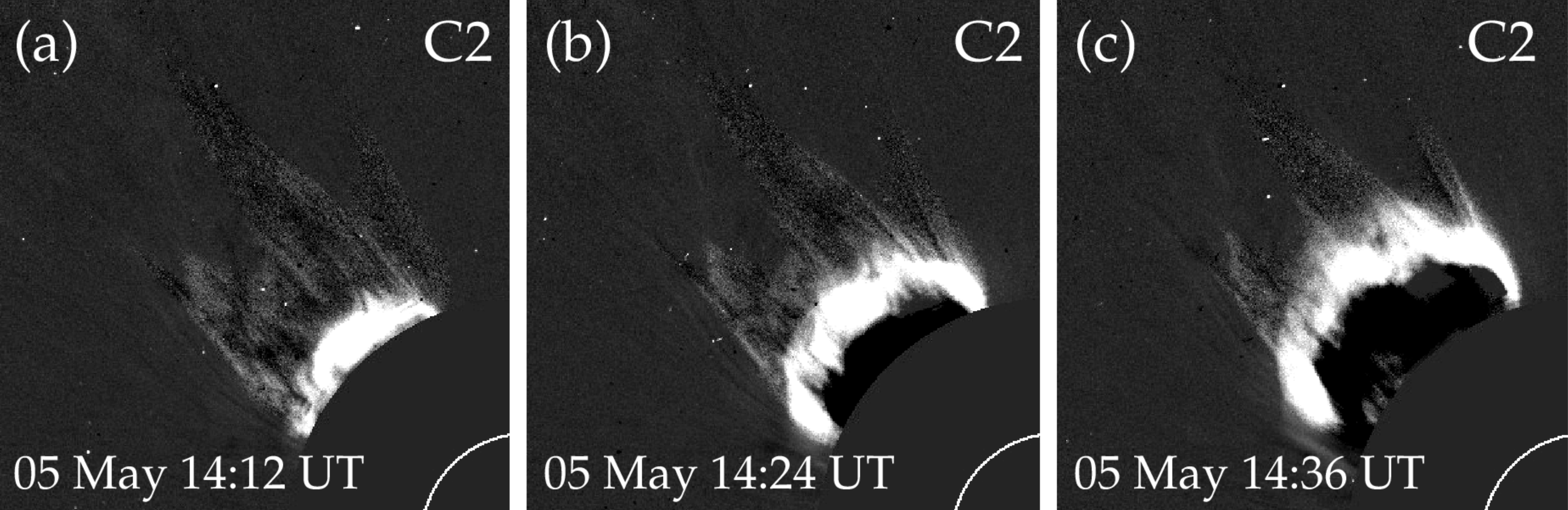}
\caption[caption ] {Running-difference images from LASCO-C2 for the CME event preceded the CME event under study. CME first appeared in the  LASCO-C2 field of view at  $\approx$14:12 UT on 05 May, 2015.}
\label{int-cme}
\end{center}
\end{figure*}

On 05 May 2015, the active region, AR2339, also produced another intense 
X-ray flare of intensity M1.2, which peaked at $\approx$13:53 UT in the 1--8 
{\AA} channel of the GOES-15 spacecraft. This event also originated at N15E75. 
However, the X-ray profile reveals it to be a short-duration event.
The onset of the CME associated with this flare event was observed at 
14:12 UT in the C2 field of view of LASCO (Figure \ref{int-cme})  and its width was $\approx$70$^{\circ}$. 
It also propagated along the position angle of $\approx$50$^{\circ}$ and 
produced Type-III and Type-II radio bursts. However, the radio signatures 
were much less intense than that of the above later CME and the Type-III burst did not cause a long-duration event. The speed of the 
Type-II shock associated with the CME was $\approx$1000 kms$^{-1}$. However, 
the speed of the CME in the LASCO C2/C3 fields 
of view was low at $\approx$350 kms$^{-1}$. This CME preceded the above 
discussed later CME at 22:24 UT (Figure \ref{lasco-diff}). Since the speed of the later CME was more 
than twice the speed of the preceding CME, in the course of time the later
one interacted with it at $\approx$50--75 {R$_{\odot}$} at about 08:00 to 10:00 UT. 
The interaction characteristics are weakly shown by the low-frequency
radio spectrum ($\leq$1 MHz part; indicated by white circles in Figure \ref{comb-spec}c) in the  time span of $\approx$10:00 to 14:00 UT \citep{gopal2001b,lahkar2010}.
The interplanetary scintillation
observations at the ORT show the turbulence and speed associated 
with the interaction phenomenon in the inner heliosphere, at distances 
outside the LASCO field of view.

\subsection{Interplanetary Scintillation Observations of CMEs}
\label{ips-obs}
In this study, interplanetary scintillation (IPS) measurements, obtained
from the ORT at 327 MHz, have been 
employed to investigate
the view of the ambient solar-wind (\textit{e.g.} Manoharan, 2012) and the 
three-dimensional evolution of the CME in the inner heliosphere  
\citep{mano2001, mano2006}. At the ORT, 
everyday routine monitoring of IPS is made on a large number of compact 
radio sources and the measurement on each source yields the estimates 
of solar-wind speed and normalised scintillation index (\textit{g}-value) at a
point closest to the solar approach of the line-of-sight to the source \citep{mano1990, mano2006}.  The normalised index (\textit{g}-value) represents the turbulence level of the solar-wind ($\Delta{N}_{e}$). At Ooty, on 06 May 2015, two scans of 
the sky were observed (Figure \ref{vel-gval}), in a distance range of 40 to 250 {R$_{\odot}$}. Each
scan included a number of sources. The IPS observation on each source 
represents the line-of-sight integration and a computer-aided tomographic
reconstruction of IPS measurements from a large number of lines-of-sight would 
be useful to remove the effect of integration \citep{jackson1997, mano2010, jackson2015}. However, in this study, we use the raw IPS data to analyse
the propagation effects of the CME and CME--CME interaction in the inner 
heliosphere.

\subsubsection{IPS Images of CME}
\label{ips-image}
Figure \ref{ips-plot} shows the snapshot scintillation images (\textit{g}-maps) of the
interplanetary medium on 06 May 2015, respectively, during 00:00--09:00 UT and
10:00--17:00 UT. {As indicated in each of the above images, the IPS measurements
have been made while the ORT was pointed, respectively, at hour angle (HA)
positions of -03h 40m and +04h 30m. Normally at the ORT, to start with, we scan the interplanetary medium at a suitable position and  then the ORT is moved ahead to point at a similar part of the heliosphere. Such observing scans provide the solar-wind measurements at similar parts of the heliosphere and can enable one to track interplanetary disturbances (\textit{e.g.} CMEs), if present, as well as change in the solar-wind conditions. In fact,  more than one scan will be useful to track the propagation characteristics of the disturbance. For example when the telescope was positioned
at the HA = -03h 40m, at the start of the scan it probed sources at
$\approx$100$^{\circ}$
solar elongation on the west side of the Sun. It is to be noted that since the
hour angle of the ORT is fixed, a radio source is observed for about two minutes at
the time of its transit at the telescope. In the subsequent time, the electronic
beam switching of the 12-beam system of the ORT allows observing sources at
different declinations and enables to probe different parts of inner heliosphere
(refer to Swarup \textit{et al.}, 1971; Manoharan \textit{et al}., 2000). As the time progresses, the
direction of the probing region moves gradually close to the Sun and then to the
portion of the heliosphere in the east side to the Sun (also refer to Figure \ref{vel-gval}).
The high sensitivity of the ORT and its beam-forming system enable the observation
of a large part of the heliosphere in about six hours of observing time. In the present case, we moved the ORT by $\approx$8h 10m in hour angle to the west (\textit{i.e.} HA = +04h 30m) and probed the heliosphere region from the west of the Sun to the east (Figure \ref{ips-plot}, right panel).}

The above images cover 500 $\times$ 500 pixels and they are smoothed by a
Gaussian of width 5 $\times$ 5 pixels. They are equivalent to
the white-light images in the sky plane projection and are useful to follow the three-dimensional
evolution of the ambient solar-wind as well as the turbulent regions
associated with the propagating disturbances in the IPS field of view of
40--250 {R$_{\odot}$}. In these images, \textit{g}-values close to unity correspond to
the ambient level of density turbulence of the solar-wind and it is
represented by the red colour code, \textit{i.e.} g$\approx$1. In contrast, the enhanced
or depleted level of density turbulence is indicated by, respectively,
\textit{g}-value $>$1 or $<$1. In the above
images, the concentric circles are of radii 50, 100, 150, and 200 {R$_{\odot}$}.

{In these ``position angle(PA)-heliocentric distance'' images, the north is at the top, \textit{i.e.} PA = 0$^\circ$. The PAs, 90$^\circ$,
180$^\circ$, and 270$^\circ$, respectively, correspond to east, south, and
west of the Sun (Figure \ref{ips-plot}, left panel).
The enhanced scintillation indicates the presence of an interplanetary CME. For
example, on the 06 May IPS image, the onset of the CME in the IPS field of view
is seen between 50 and 100 {R$_{\odot}$} and it corresponds to a time period of
$\approx$04:00--06:00 UT (refer to Figure \ref{vel-gval}, which shows the time-series of estimates of solar-wind speed and \textit{g}-value). It is noted that the
disturbance associated with the CME has expanded considerably in the IPS
field of view and the comparison between these images clearly shows the
propagation of the CME in the eastern direction with respect to the Sun.} The
time-series analysis of the Ooty observations reveals a better understanding
of the CME propagation as well as interaction between the above discussed CMEs.

\subsubsection{IPS Time Series}
\label{ips-time}
\begin{figure}
\begin{center}
\includegraphics[width=12cm,angle=180,clip=true]{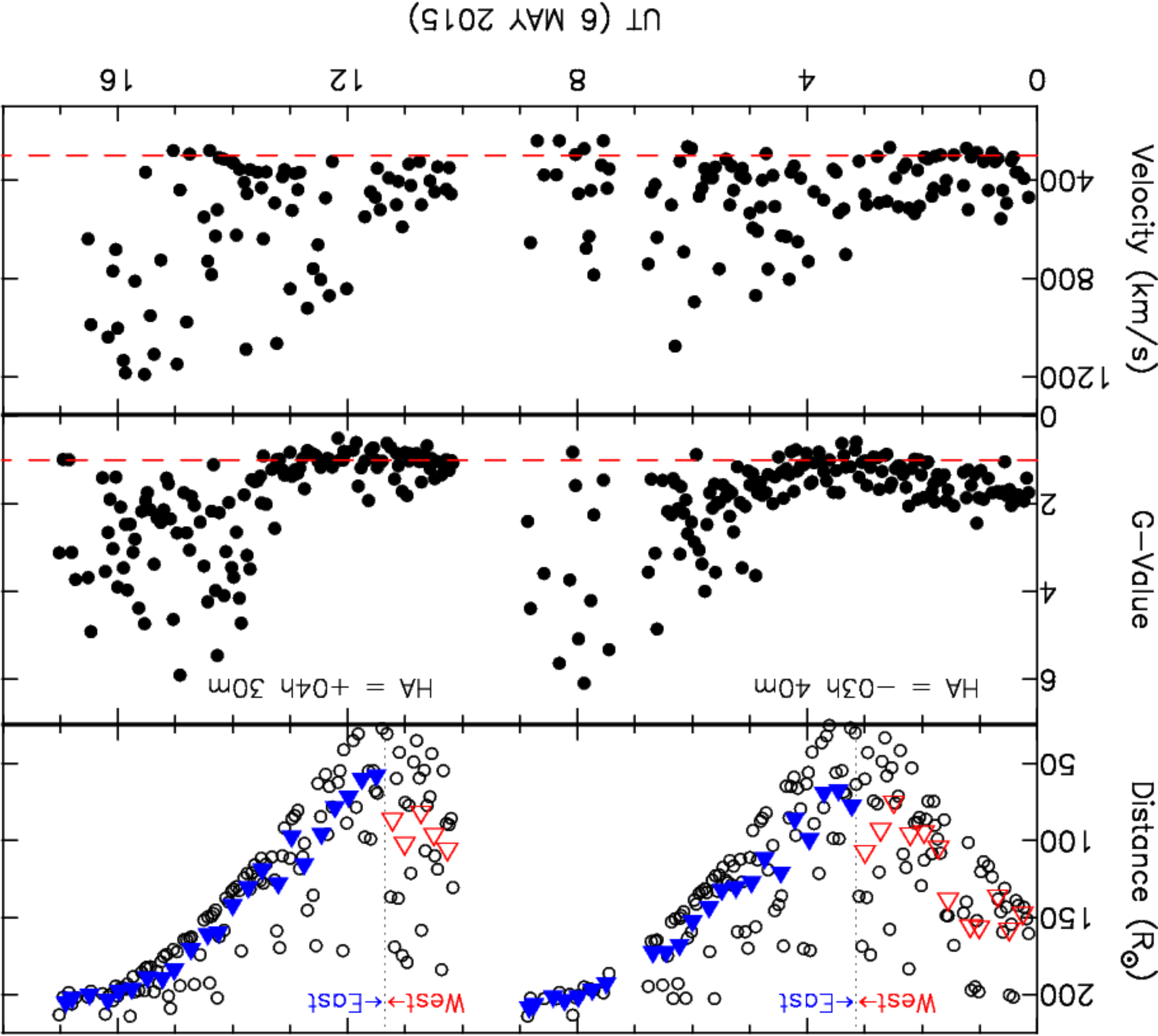}
\caption[caption]{Time series of IPS observations on 06 May 2015. The bottom and middle panels show velocity and \textit{g}-value plots, respectively. {The horizontal dashed lines on speed and \textit{g}-value plots respectively indicate the background solar-wind speed of $\approx$300 kms$^{-1}$ and turbulence level of g$\approx$1. In the top panel, the circle symbol represents the heliocentric distance to the line of sight to the radio source and triangles (open and filled) represent the distance average over $\approx$20 minutes of observation. The separation between west and east sides of the heliosphere probed with respect to Sun is shown by a vertical-dotted line.}}
\label{vel-gval}
\end{center}
\end{figure}

\begin{figure*}
\begin{center}
\includegraphics[width=6.0 cm,angle=0,clip=true]{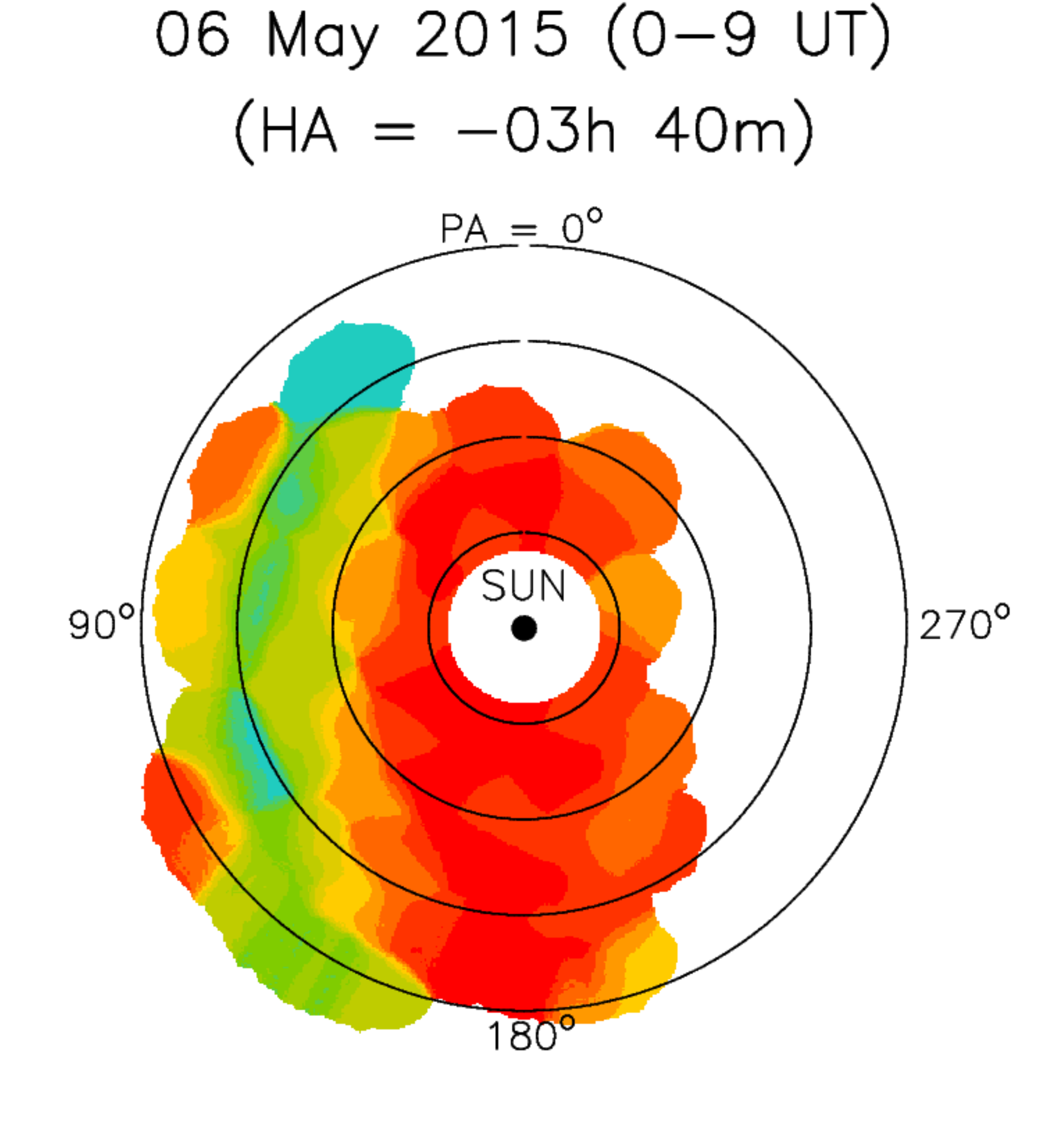}
\includegraphics[width=6.0 cm,angle=0,clip=true]{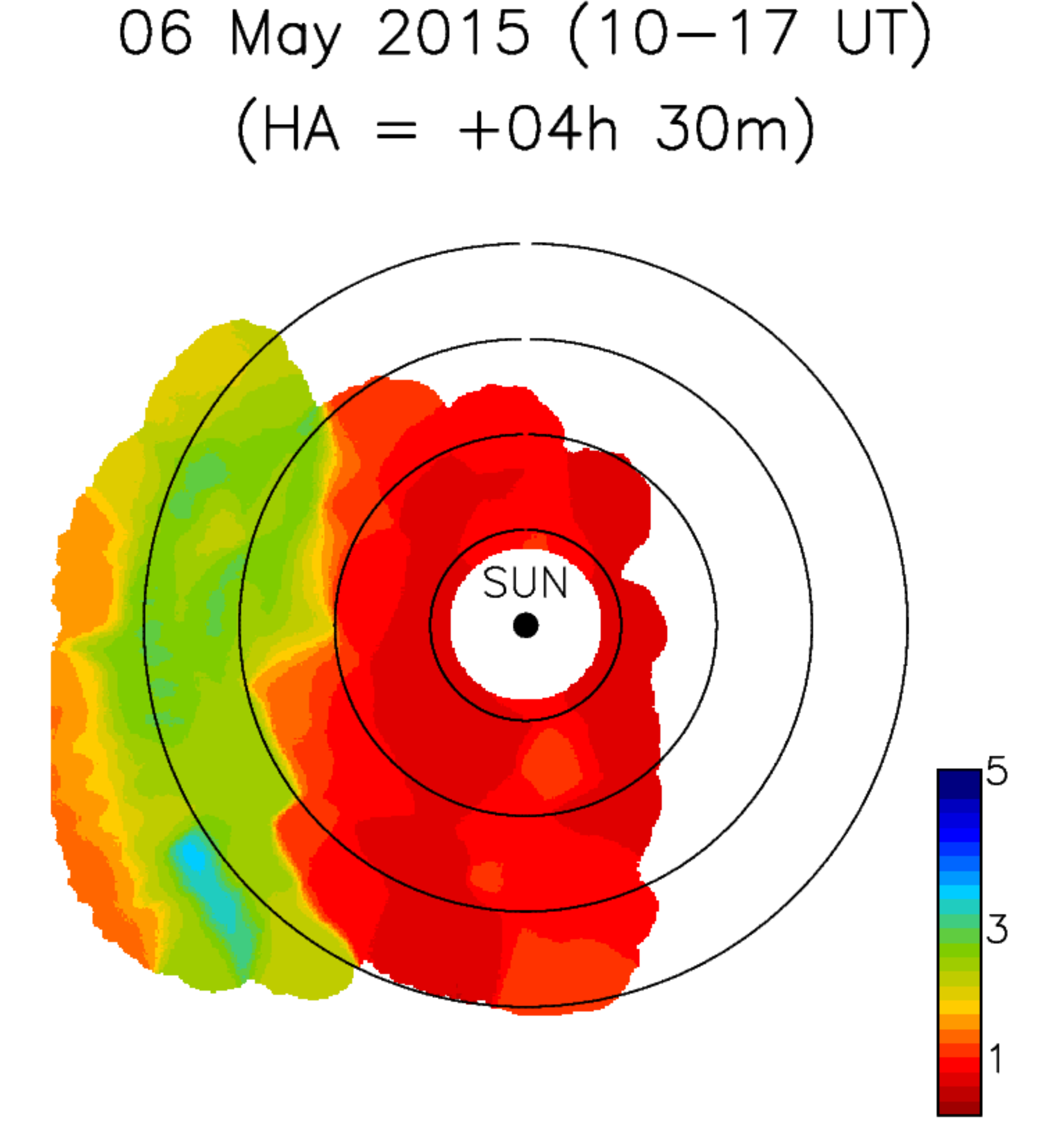}
\caption[caption ]{Interplanetary scintillation images observed with the ORT on 06 May 2015. In these ``position angle--distance'' diagrams, Sun is at the centre and concentric circles are of radii 50, 100, 150, 200 {R$_{\odot}$}. The colour scale shows the normalised scintillation index (\textit{g}-value). Time increases from the right-hand side of the image to the left-hand side of the image. {The indicated hour angle (HA) positions correspond to the ORT pointing directions. The position angles (PAs), 0$^\circ$, 90$^\circ$,
180$^\circ$, and 270$^\circ$, respectively, correspond to north, east, south, and
west of the Sun. }}
\label{ips-plot}
\end{center}
\end{figure*}

Time series of speed and \textit{g}-value estimates obtained from the Ooty IPS
measurements on 06 May 2015 are shown in Figure \ref{vel-gval}, which
includes two scans of observations, respectively, 00:00--09:00 UT and
10:00--17:00 UT. Each point on the plot represents the observation from
a scintillating radio source. {As discussed in the previous section, the
above two scans were observed, respectively, at hour angles -03h 40m and
+04h 30m. In each scan, to start with the western heliospheric region with
respect to the Sun is probed and as the time progressed, the probing direction  moves close to the Sun and later  to the east of
the Sun. On each scan of the ``distance--time'' plot ({top} panel in
Figure \ref{vel-gval}), the heliocentric distance of an observed source is plotted with
an { open circle} symbol and the typical separation between the west and
east sides of the heliosphere probed with respect to Sun is shown by a vertical
dotted line. This plot also includes the average heliocentric distance covered
over an observing time of $\approx$20 minutes and are shown by { open-triangle}
and { filled-triangle} symbols respectively.}

{In Figure \ref{vel-gval}, the dashed lines on the speed and \textit{g}-value plots respectively
indicate the background solar-wind speed of $\approx$300 kms$^{-1}$ and turbulence
level of g$\approx$1. The estimation of the \textit{g}-value can be obtained from most
of the scintillating radio sources and the \textit{g}-plot includes observations of
about 325 radio sources, covering a wide range of distances in the heliosphere.
In contrast, the speed estimate from an IPS observation is limited to a temporal
spectrum of high signal-to-noise (\textit{i.e.} S/N $\geq$ 15 dB) and the speed plot
has been made from $\approx$200 data points of significantly high signal-to-noise
ratio observations. }

In Figure \ref{vel-gval}, at the eastern side of the Sun, the heliocentric
distance increases with the observing time. The first scan, observed during
$\approx$00:00--09:00 UT, shows the following solar-wind conditions: {i) the
ambient/background solar-wind is probed before the onset of the CME(s) in
the IPS field of view; ii) after the onset of the CME in the IPS field of
view, most of the lines of sight of the radio sources pass through different
parts of the propagating CME structure and their \textit{g}-values provide the typical
size of the interplanetary CME structure; iii) only a few radio sources point
outside the CME structure and their \textit{g}-values lie close to the background line;
iv) at the onset time of the CME, the lines of sight pass through the CME structure
in the distance range of $\approx$50 and 100 {R$_{\odot}$}, which is observed as a gradual
rise in \textit{g}-value in the time period of $\approx$04:00--07:00 UT; v) timings and distances
involved in these observations also indicate that the CME is propelled fast at 
speed $>$800 kms$^{-1}$; vi) however, the turbulence level associated with 
it is less than that observed at $\approx$08:00 UT; vii) at distances $>$100 {R$_{\odot}$} 
(\textit{i.e.} at times $>$06:00 to 09:00 UT), the solar-wind region between CMEs, and the 
preceding CME are probed; viii) around 08:00 UT, the speed plot indicates the low speed associated with the preceding CME.}

As indicated by the speeds of these CMEs in the LASCO field of view as
well as by radio signatures, the CME--CME interaction is likely to happen
around 08:00--10:00 UT, at a height of about 50--75 {R$_{\odot}$}. The second IPS scan
of the interplanetary medium pictures the status of the solar-wind just
after the interaction. The condition of the solar-wind after the interaction
of CMEs is, i) the overall turbulence level
is high, ii) the average speed of the solar-wind is well above the
background solar-wind, iii) as shown by the IPS images, the intense
part of the turbulence is wide and it includes both the CMEs.
In the second scan, the high speed points are also observed before the
peaking of \textit{g}-values, \textit{i.e.} behind the CMEs.

Another important point to be noted is that as revealed by the IPS 
observations, the CME before and after interaction has travelled well 
above the speed of the background solar-wind. In fact, the effective 
drag force experienced by a CME is largely determined by the difference 
in speeds between the CME and the ambient solar-wind and it is proportional
to $|$V$_{CME}$ - V$_{ambient}$$|$$^{2}$ \citep{mano2006}. In the present case, the speed of
the CME is $>$800 kms$^{-1}$, which would have suffered the reduction of
speed due to the interaction of the CME with the ambient solar-wind of speed 
$\approx$300--350 kms$^{-1}$. Since the CME seems to continue to propagate 
at a speed higher than the ambient solar-wind, it is possible that the 
internal magnetic energy  associated with the filament (or flux rope) 
supports to propel the CME structure out into the solar-wind \citep{chen1997,mano2011}.

\section{Discussion and Summary}
\label{disc-sum}

The propagation effects of the (X2.7 related) fast CME of 05 May 2015 
have been analysed using a combination of data from EUV, radio, 
white-light, and IPS observations in the Sun to 1-AU distance range. 
The filament oscillation is first seen in the AIA 171 {\AA} images at 
22:05 UT, just prior to the onset of the eruption. The fast eruption 
of filament is observed with a velocity of $\approx$1300 kms$^{-1}$ in 
the time range of 22:10--22:14 UT at heights below 1.5 {R$_{\odot}$} and 
in this period, the rate of expansion of the CME in the lateral 
direction (\textit{i.e.} perpendicular to the direction of propagation) is confined and it is much smaller than the eruption speed. 
The intense Type-III radio burst, along the open field lines above the
eruption site, extends from the low corona (\textit{i.e.} at frequency $\approx$500 
MHz) all the way into the interplanetary medium (frequency $\approx$100 
kHz) and it suggests the possible acceleration of particle by post-CME 
reconnection \citep{mck2001,sheeley2004}. 

The speed of the CME-driven disturbances, from the white-light 
images and the shock shown by the metric Type-II bursts, increases 
with height, peaking at $\approx$1600 kms$^{-1}$ at a {solar-height} of $\approx$6 
{R$_{\odot}$} and then the speed stabilises at $\approx$840 kms$^{-1}$ at greater
heights. The LASCO images show the quick expansion of the filament 
enclosed within the CME structure. This suggests that the CME expands
to maintain the pressure balance with the low-speed ambient solar-wind \citep{mano2000}.

The interplanetary-scintillation observations show the large-scale
propagation of the CME through the interplanetary medium, its expanding 
structure and distribution in the inner heliosphere. Even at heliocentric
distances $\geq$60 {R$_{\odot}$}, the CME-driven disturbances move with speed
in excess of 800 kms$^{-1}$, which shows that the effective drag imposed 
by the low-speed background solar-wind seems to be ineffective.
It is likely that the internal magnetic energy in the filament aids the 
expansion and propagation \citep{moore1992, chen1997b, demoulin1998}. {The density depletion caused
by a preceding CME could also aid the propagation of the later CME. However, in the present case, the prior CME could have started with a high initial speed, and in the LASCO field of view it  likely slowed to a speed close to the ambient solar-wind speed.}

The IPS observations on a large number of sources allow us to look at
the interaction between the fast CME and a slow-moving preceding CME,
at the heliocentric-distance range of $\approx$50--75 {R$_{\odot}$}. It has
been shown that the interaction between CMEs leads to slowing down of
the fast CME \citep{mano2004}. As the
result of interaction, the turbulence level contributed by both the CMEs
has increased significantly. Since at these heliocentric distances,
the level of turbulence is related to $\Delta N_e$ of the solar-wind,
the density $\left[ N_e\right]$ within the merged or interacted region is also
expected to increase considerable. Although it is not straightforward to infer the factor of increase of the solar-wind density, it is inferred that the
combined effect of the CME structure and propagation characteristics
at further large distances would considerably alter the arrival of
the fast CME at 1 AU.

\begin{acknowledgements}
 We thank the observing and engineering staff of the Radio Astronomy 
Centre (RAC) for help in making the IPS observations. The RAC is 
run by the National Centre for Radio Astrophysics of the Tata 
Institute of Fundamental Research. We acknowledge the LASCO
images from the SOHO mission, which is a project of international 
cooperation between ESA and NASA. The authors also thank  Solar 
Dynamics Observatory (SDO) for providing the high resolution EUV
images. We acknowledge the teams of GOES spacecraft, ISTP \textit{Wind}/WAVES 
and Hiraiso Radio Spectrograph (HiRAS\footnote{hirweb.nict.go.jp}). We are thankful to the World Data Centre (WDC) of Space Weather Services, Bureau of Meteorology of Australia for the analysis of Culgoora\footnote{www.sws.bom.gov.au/Solar/} data. PROBA-2/SWAP is a project of the Centre Spatial de Liege and the Royal
Observatory of Belgium funded by the Belgian Federal Science Policy
Office (BELSPO). We also like to thank MA Krishnakumar for reading the manuscript. We thank the referee for the useful comments.
\end{acknowledgements}

\section*{Disclosure}
Abhishek Johri and P.K. Manoharan do not have any conflicts of interest.



\end{article} 
\end{document}